\documentstyle[pre,aps,epsf]{revtex}
\def\be{\begin{equation}}
\def\ee{\end{equation}}
\draft
\title{Crossover properties from random percolation to
frustrated percolation}
\author{Luigi Cannavacciuolo, Antonio de Candia, 
and Antonio Coniglio}
\address{Dipartimento di Scienze Fisiche, 
Mostra d'Oltremare, Pad. 19, 80125 Napoli, Italy}
\address{INFM, Sezione di Napoli}
\begin{document}
\twocolumn
\wideabs{%
\maketitle
\begin{abstract}
We investigate the crossover properties of the 
frustrated percolation model on a two-dimensional square
lattice, with asymmetric distribution of 
ferromagnetic and antiferromagnetic interactions. We
determine the critical exponents $\nu$, $\gamma$, and $\beta$ 
of the percolation transition of the model,
for various values of the density of
antiferromagnetic interactions $\pi$ in the range $0\leq\pi\leq 0.5$.
Our data are consistent with the existence of a 
crossover from random percolation behavior for $\pi=0$, to frustrated 
percolation behavior, characterized by the critical exponents 
of the ferromagnetic $1/2$-state Potts model, as soon as $\pi>0$.
\end{abstract}
}
%
%
\section{Introduction}
The cluster approach was introduced by Kasteleyn and Fortuin (KF)
\cite{ref_kasteleyn}
and Coniglio and Klein (CK) \cite{ref_cklein} in ferromagnetic spin systems.
In the Coniglio-Klein approach, one puts a bond between two nearest 
neighbor (NN) spins with probability $p=1-e^{-2\beta J}$
if the spins are parallel, where $J$ is
the spin interaction and $\beta=1/k_BT$, and with probability zero if they 
are antiparallel.
In this way
a bond configuration $C$ on the lattice has a statistical weight
\be
W(C)=e^{\mu b(C)} q^{N(C)},
\ee
where $\mu=\log\left(\frac{p}{1-p}\right)=\log(e^{2\beta J}-1)$ 
is the chemical potential of the bonds,
$b(C)$ is the number of bonds and $N(C)$ the number of clusters 
of the configuration $C$, and $q$ is the multiplicity of the spins
($q=2$ for Ising spins).
This defines a percolation model, in which clusters percolate 
at the ferromagnetic
critical point, with the same critical indices of the original spin model.

The KF-CK approach has been extended also to
frustrated spin models, as for example spin glasses \cite{ref_cdiliberto}.
In these models the disorder is produced by a quenched
distribution of antiferromagnetic ($-J$) and ferromagnetic ($J$)
interactions on the lattice. 
Just like in the ferromagnetic case, one puts a bond between two 
NN spins if they satisfy the interaction, with a probability
$p=1-e^{-2\beta J}$. The main difference here is that, due to frustration,
the spins cannot satisfy simultaneously all the interactions on the
lattice. More specifically, if a closed path on the lattice contains
an odd number of antiferromagnetic interactions, not all the interactions
belonging to the path can be satisfied simultaneously. 
Such a path is called ``frustrated loop'', and since 
bonds can be put only between spins that satisfy the interactions,
a frustrated loop cannot be completely occupied by bonds.
Therefore, the statistical weight of a
bond configuration $C$ will be now
\be
W(C)=\left\{\begin{array}{ll}
e^{\mu b(C)} q^{N(C)}\quad&
\text{if $C$ is not frustrated,}\\
0&\text{if $C$ is frustrated,}\\
\end{array}\right.
\ee 
where $C$ is said frustrated if it contains one or more frustrated loops
completely occupied by bonds.

It has been shown by renormalization
group methods on a hierarchical lattice
\cite{ref_cpezzella} that
this model exhibits two phase transitions, for every value of the
multiplicity $q$ of the spins.
The first transition, at a temperature $T_{SG}(q)$, is in the universality
class of the Ising SG transition, while
the other transition,  at a temperature $T_p(q)>T_{SG}(q)$, is a
percolation transition in the universality class of the ferromagnetic
$q/2$-state Potts model \cite{ref_wu}.

The frustrated percolation  model has proven to be a 
suitable model to the study of complex systems,
such as spin glasses \cite{ref_sg}, glasses \cite{ref_glass}
and in general all those systems in which
connectivity and frustration play a fundamental role,
(for a review see \cite{ref_review}).

The aim of the present paper is the study of the
percolation transition of the model with $q=1$,
the ``bond frustrated percolation model'',
for a variable density $\pi$ of antiferromagnetic interactions 
in the interval $0\leq\pi\leq 0.5$.
We determine the critical probability 
$p_c(\pi)$ and the critical exponents $\nu(\pi)$, $\beta(\pi)$, 
and $\gamma(\pi)$,  by performing Monte Carlo simulations on
lattices of different size, and using scaling laws 
(see Sect. \ref{sec_results}).
Finally the results are compared
with the theoretical predictions for the two extreme cases, 
the pure ferromagnetic case $\pi=0$,
that corresponds to random bond percolation 
($1/\nu=0.75$, $\beta/\nu=0.1042$, $\gamma/\nu=1.7917$),
and the symmetric case $\pi=0.5$, that corresponds to the $1/2$-state
Potts model
($1/\nu=0.5611$, $\beta/\nu=0.08276$, $\gamma/\nu=1.8346$).  
\section{Definition of the frustrated percolation model}
\label{sec_model}
The bond frustrated percolation model can be defined in the following way.
Consider a two-dimensional square lattice, 
with NN interactions between sites.
These interactions can be ferromagnetic with probability $1-\pi$ and 
antiferromagnetic with probability $\pi$. The distribution of interactions is 
quenched, so it is set at the beginning and does not evolve with time in the 
dynamics of the system. 
Each edge of the lattice, connecting a pair of NN sites, can be connected by
a bond or not, and the state of the system is completely specified
by the bond configuration.

We give to a bond configuration $C$ a statistical weight
\be
W(C)=\left\{
\begin{array}{ll}
e^{\mu b(C)} 
&\quad\text{if $C$ is not frustrated,}\\
0&\quad\text{if $C$ is frustrated,}
\end{array}\right.
\ee
where $\mu=\log\left(\frac{p}{1-p}\right)$, $p$ is a probability that is
connected to the temperature via the relation $p=1-e^{-2\beta J}$,
and $b(C)$ is the number of bonds of the configuration $C$.

The presence of frustration 
induces a complex behavior in both the static and dynamic properties
of the system,
with the presence of many metastable states, and
high free energy barriers separating them. 
For $\pi=0$ the model coincides with random percolation.
For $\pi=1/2$, that is
equal density of ferromagnetic and antiferromagnetic interactions,
the system is expected to have two phase transitions \cite{ref_cpezzella}. 
The first 
at lower temperature  in the same universality
class of the Ising spin glass transition,
which in two dimensions is at $T=0$,
that is at $p=1$. Below this temperature
the system is frozen, ergodicity is broken and the system
remains trapped in a finite region of phase space. The second 
transition at a higher temperature (lower probability),
is the percolation transition of the cluster of bonds,  
and belongs to a different universality class, namely 
that of the ferromagnetic $1/2$-state Potts model.

We will study the percolation transition as a function of the
density of antiferromagnetic interactions $\pi$. 
We will see that a very small amount 
of antiferromagnetic interactions  is already sufficient to
change the universality class of the transition.  
\section{Monte Carlo results}
\label{sec_results}
We have studied the percolation transition of the model
on a 2D square lattice with periodic
boundary conditions, for different values of $\pi$
($0\leq\pi\leq0.5$) and for different lattice sizes $L$
($L=16,24,32,40$, occasionally larger). For each size of the
lattice a configuration of interactions
is preliminary produced by setting randomly on the
lattice a number $N_a$ of
antiferromagnetic interactions and $N_f=2L^{2}-N_a$ of
ferromagnetic interactions. The corresponding density of antiferromagnetic
interactions is given by $\pi=N_a/2L^2$.

We have determined the
critical probability of percolation $p_{c}(\pi)$ and the
critical exponents $\nu(\pi)$, $\gamma(\pi)$,
and $\beta(\pi)$, by performing Monte Carlo simulations and
using scaling laws. In order to generate bonds
configurations with the appropriate weights we used an
algorithm devised by Sweeny \cite{ref_sweeny}, 
suitably modified
to treat the frustration occurrence \cite{ref_antonio}. For
each $\pi$ the quantities of interest were averaged on
many configurations of interactions.

For each value of $\pi$ the critical probability of
percolation has been determined using the following
scaling law for the probability $P_\infty$ of a spanning
cluster being present on the lattice \cite{ref_stauffer}
\be
P_\infty(L,p)=\widetilde{P}_\infty[L^{1/\nu}(p-p_{c})],
\quad\text{for}\> L\to\infty, p\to p_c.
\label{eq_Piscal}
\ee
Therefore $p_{c}$ is found to be the point where the
curves of $P_\infty$ as a function of $p$, with different lattice sizes, 
intersect.

\begin{figure}
\begin{center}
\mbox{\epsfysize=8cm\epsfbox{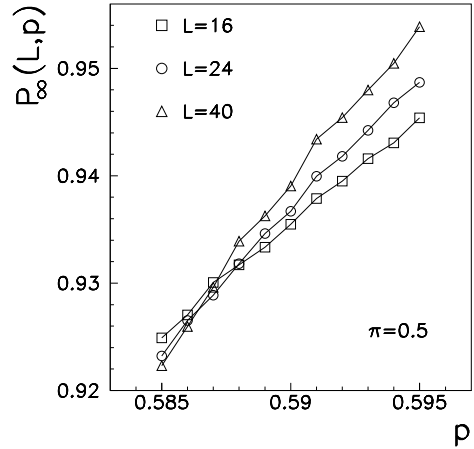}}
\end{center}
\caption{Probability of percolation $P_\infty$,
as a function of the probability $p$, for lattice sizes $L=16,
24, 40$ and for $\pi = 0.5$. The critical probability
$p_{c}$ is found to be the intersection point of the
different curves.}
\label{fig_pc_05}
\end{figure}

\begin{figure}
\begin{center}
\mbox{\epsfysize=8cm\epsfbox{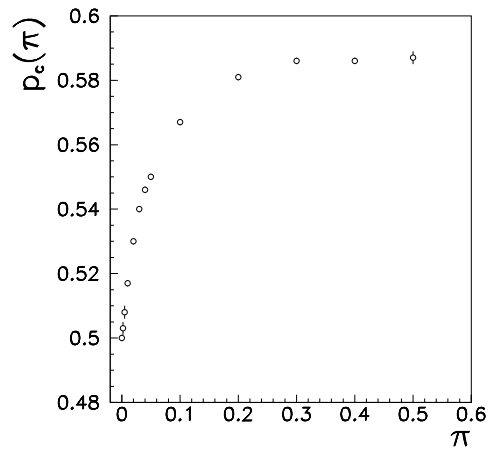}}
\end{center}
\caption{The critical probability of percolation $p_{c}$
as a function of the density of antiferromagnetic interactions
$\pi$. The value $p_{c}=0.5$ found for $\pi=0$ is in
agreement with the value of the random percolation.}
\label{fig_pc}
\end{figure}

In Fig. \ref{fig_pc_05} are shown the curves $P_\infty(L,p)$
for $\pi=0.5$,
and in Fig. \ref{fig_pc} the values obtained for
$p_{c}(\pi)$. The intersection point is clearly singled
out. This procedure enabled us to determine $p_{c}$ with
an indetermination of $\pm 0.001$, or in the worst cases
of $\pm 0.002$.

The simulations were performed on lattices of size $L=16$, 24, 32,
40, for each $L$ about $500$ MC steps were produced to
thermalize the system whereas $20,000$-$30,000$ MC steps
were used to average. Moreover for each value of $\pi$ we
have averaged on a number $n$ of configurations of
interactions ranging from $n=80$ for $L=16$ to $n=30$ for
$L=40$.

\begin{figure}
\begin{center}
\mbox{\epsfysize=8cm\epsfbox{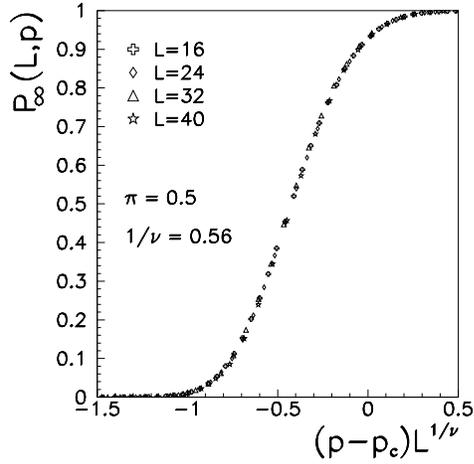}}
\end{center}
\caption{Scaling behavior of the probability of percolation $P_\infty$,
for $\pi=0.5$. The exponent $1/\nu$ is found as the value
which gives the best data collapse. The value $1/\nu=0.56$ found
is in agreement with the value of the 1/2-state Potts model.}
\label{fig_ni_05}
\end{figure}

\begin{figure}
\begin{center}
\mbox{\epsfysize=8cm\epsfbox{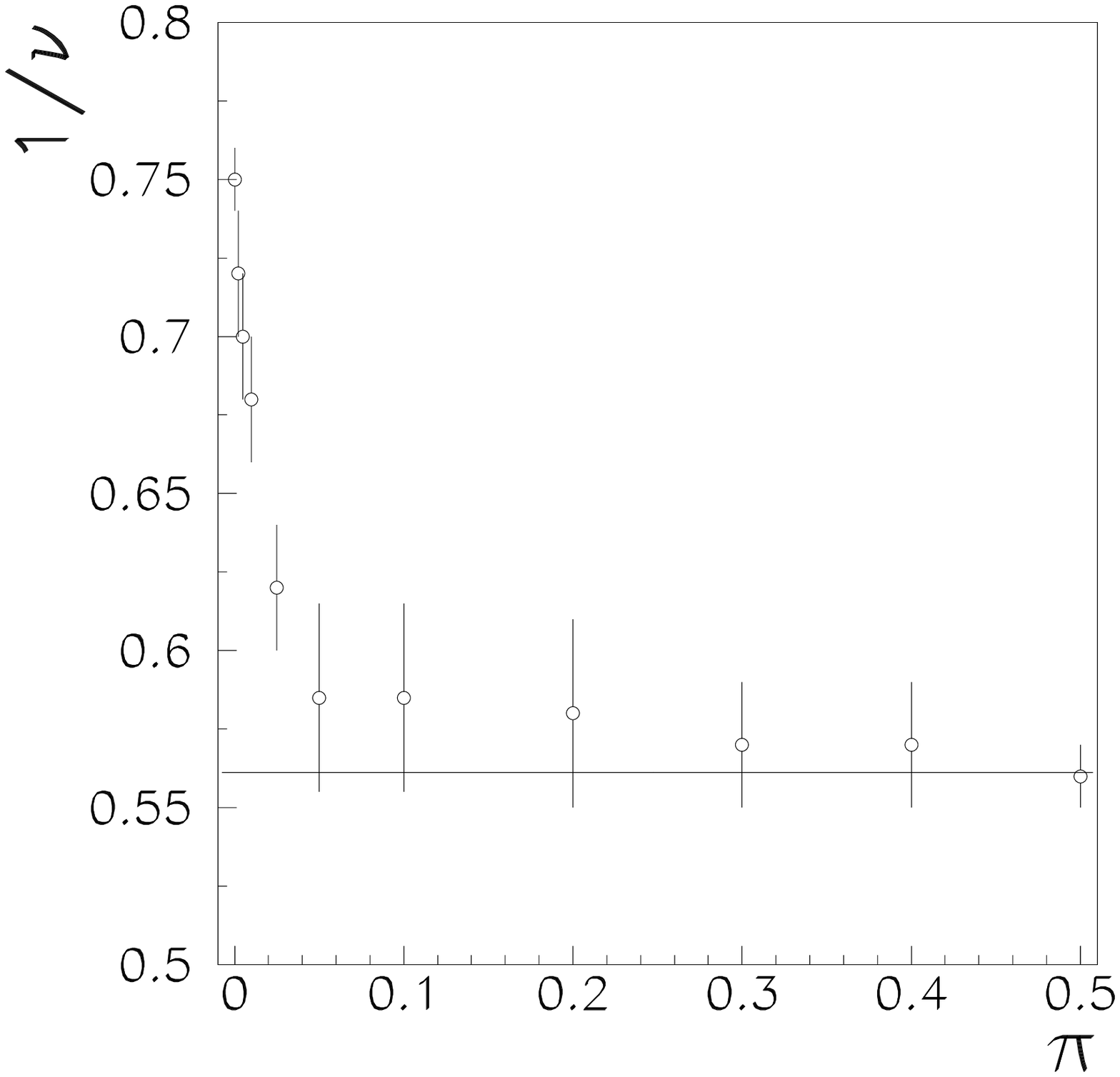}}
\end{center}
\caption{Plot of the exponent $1/\nu$ 
in function of the density of antiferromagnetic interactions
$\pi$.
The horizontal straight line represents the
value of $1/\nu$ for the frustrated percolation.}
\label{fig_ni}
\end{figure}

The scaling law in Eq. (\ref{eq_Piscal}) 
enables us to get the value of the exponent
$1/\nu$ as well, once $p_{c}$ has been determined, by
choosing the value which gives the best data collapse of
the curves (see Fig. \ref{fig_ni_05}). 
In Fig. \ref{fig_ni} we give the
values of $1/\nu(\pi)$ obtained. As $\pi$ increases from
zero to positive values, a sudden change in the universality class of
the transition can be seen. A crossover region extending
from $\pi=0$ to $\pi \simeq 0.05 $ occurs. From this
point the value of $\nu$ can be considered in
agreement with the predicted value for the $1/2$-state
Potts model ($1/\nu=0.56$,
marked by the horizontal straight line).

The errors $\Delta(1/\nu)$ and $\Delta p_c$ 
were computed as the amplitudes of the intervals in the values 
of $1/\nu$ and $p_c$ for which a good data collapse
was obtained.
We remark that these errors do not take into account the
finiteness of the system. Thus the values for the
critical exponents must be regarded as effective values
which would give correct results only in the asyntotic
limit ($L\to \infty$).

The mean cluster size $\chi$ is defined as
\be
\chi=\frac{1}{V}\sum s^2 n_s,
\ee
where $V=L^2$ is the number of sites on the lattice, 
$n_s$ is the number of 
clusters of size $s$, and the sum extends over the cluster sizes $s$.
In the thermodynamic limit $L\to\infty$ the mean cluster
size diverges as $|p-p_c|^{-\gamma}$, when the probability $p$ approaches 
its critical value. For finite systems, $\chi$ obeys a finite size scaling
\cite{ref_stauffer}
\be
\chi(L,p) = L^{\gamma/\nu} \widetilde{\chi} [L^{1/\nu}(p-p_{c})]
\quad\text{for}\> L\to\infty, p\to p_c.
\ee
The density of the largest cluster $\rho_{\infty}$ plays the role
of the order parameter in the system, being zero for $p<p_c$ in the 
thermodynamic limit, and $\rho_{\infty}\propto (p-p_c)^\beta$ 
for $p>p_c$.
This quantity as well obeys a finite size scaling 
\be
\rho_{\infty}(L,p) = L^{-\beta/\nu} 
\widetilde{\rho}_\infty[L^{1/\nu}(p-p_{c})]
\quad\text{for}\> L\to\infty, p\to p_c.
\ee

\begin{figure}
\begin{center}
\mbox{\epsfysize=8cm\epsfbox{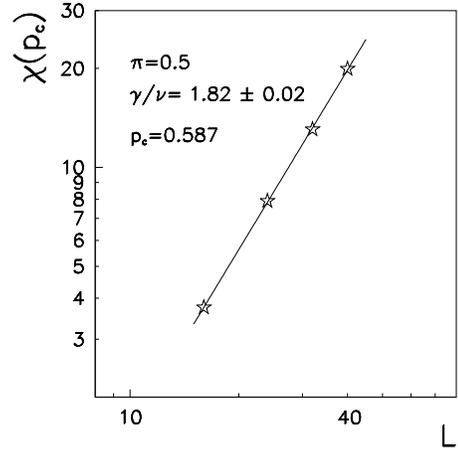}}
\end{center}
\caption{Log-log plot of the mean cluster size $\chi$ in function of
$L$, at $p_{c}$ and for $\pi=0.5$, and linear fit of the data.
The critical exponent $\gamma/\nu$ is found to be the
slope of the straight line.}
\label{fig_gamma_05}
\end{figure}

\begin{figure}
\begin{center}
\mbox{\epsfysize=8cm\epsfbox{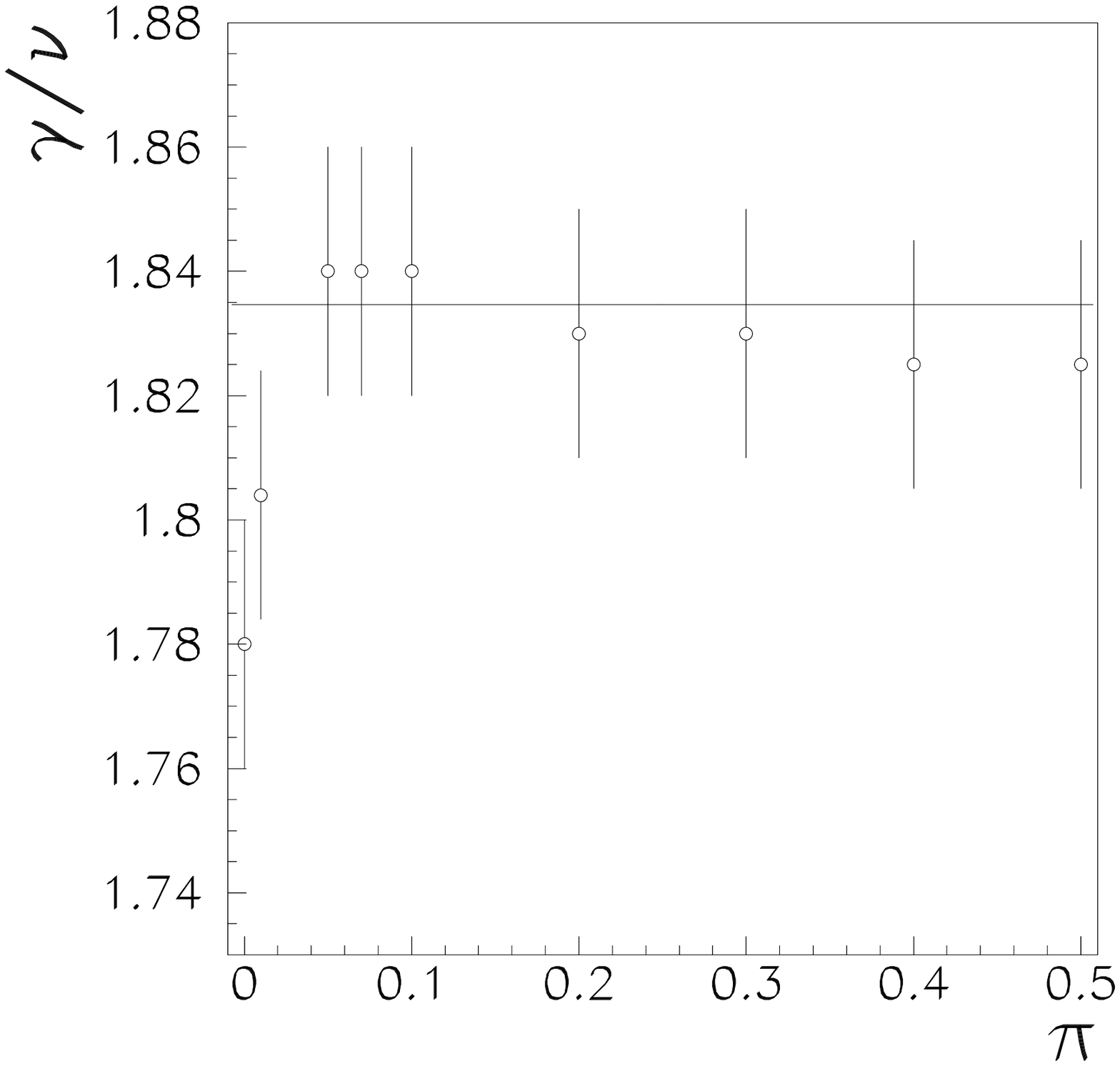}}
\end{center}
\caption{Plot of the exponent $\gamma/\nu$ 
in function of the density of antiferromagnetic interactions $\pi$.
The horizontal straight line represents the
value of $\gamma/\nu$ for the frustrated percolation.}
\label{fig_gamma}
\end{figure}

\begin{figure}
\begin{center}
\mbox{\epsfysize=8cm\epsfbox{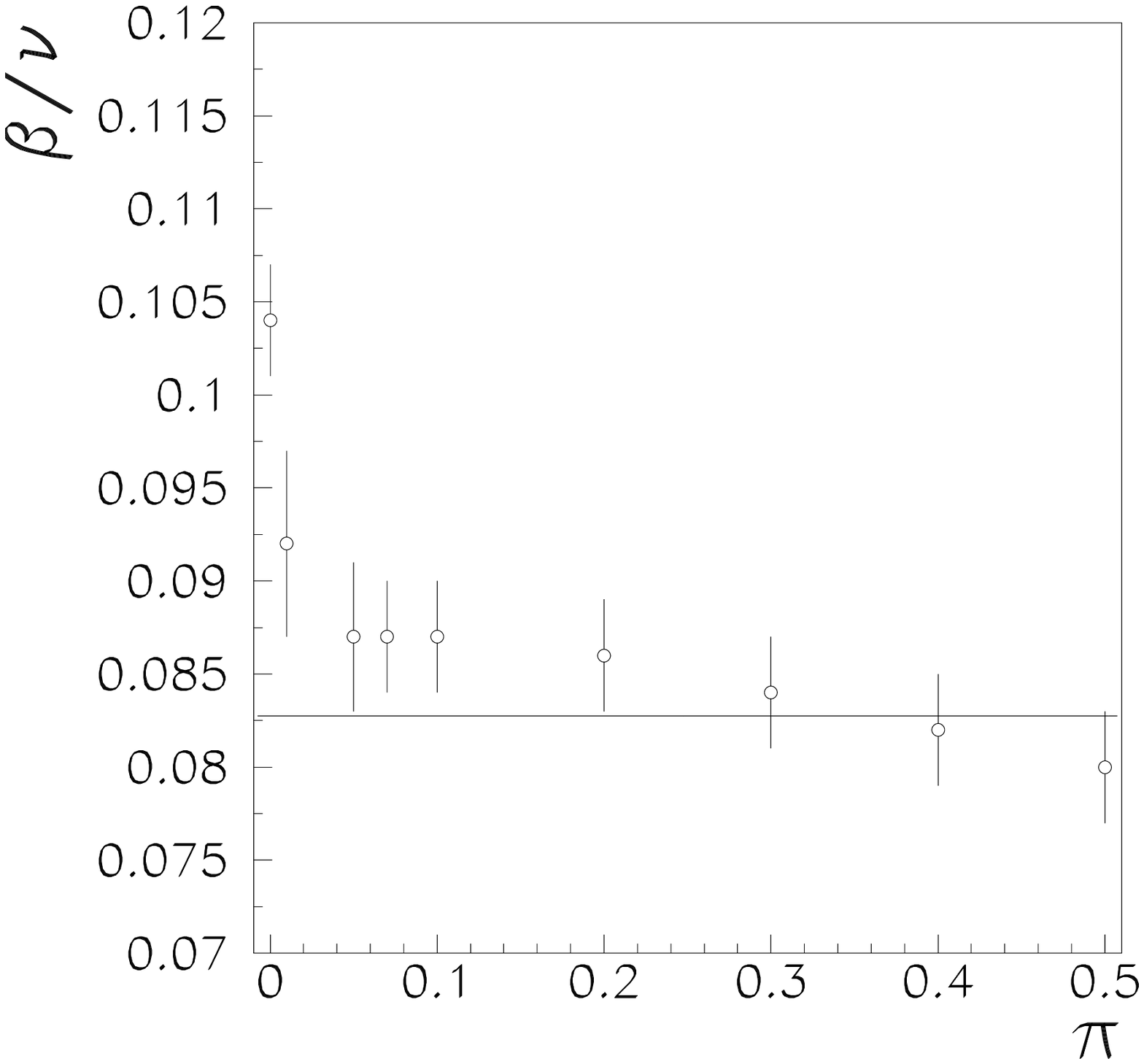}}
\end{center}
\caption{Plot of the exponent $\beta/\nu$ 
in function of the density of antiferromagnetic interactions $\pi$.
The horizontal straight line represents the
value of $\beta/\nu$ for the frustrated percolation.}
\label{fig_beta}
\end{figure}

Therefore simulating the system at the computed value of $p_{c}$
enables us to get $\gamma$ and $\beta$ from a
log-log plot of the relations
$\chi (L,p=p_c) \propto L^{\gamma/\nu}$, 
$\rho_{\infty}(L,p=p_c) \propto L^{-\beta/\nu}$.
In Fig. \ref{fig_gamma_05} one such plot is shown, for $\pi = 0$,
while in Fig. \ref{fig_gamma} and Fig. \ref{fig_beta} 
are shown the plots of
$\gamma/\nu$ and $\beta/\nu$ obtained, for different values of $\pi$.
The horizontal straight lines mark the predicted values of the
exponents, $\gamma/\nu=1.83$ and $\beta/\nu=0.083$, corresponding
to the ferromagnetic $1/2$-state Potts model. We see that for
$\pi>0.05$ the computed value of the exponents is in good agreement
with the prediction.

These results are consistent with the picture that there are two universality 
classes: random percolation at $\pi=0$, and frustrated percolation for 
$\pi>0$. Data at our disposal cannot exclude however the possibility that the 
random percolation behavior extends from $\pi=0$ to a value $\pi^\ast$ smaller 
than $0.05$.
\section{Conclusions}
We have investigated the percolation transition of the
asymmetric frustrated percolation model in two dimensions
by using Monte Carlo simulation.

From the analysis of critical exponents $\nu(\pi)$,
$\gamma(\pi)$, $\beta(\pi)$, in the interval $0\leq \pi \leq 0.5$, it
seems reasonable to assume that a very small concentration
($\pi \simeq 0.05$) of antiferromagnetic interactions is already
sufficient to produce the change in the universality
class of the transition. This means that the effects of
disorder and frustration are important even for such
small values of $\pi$.

Our results are consistent with the existence of a sharp crossover 
from random percolation for $\pi=0$, to frustrated percolation,
characterized by the exponents of the ferromagnetic $1/2$-state Potts 
model, as soon as $\pi>0$.
However we cannot rule out numerically the presence of a tricritical 
point, at low values of $\pi$, dividing random percolation from
frustrated percolation exponents.
%
%
%
%

%
%
%
%
\end{document}